\documentclass[conference,10pt]{IEEEtran}
\IEEEoverridecommandlockouts
\usepackage{cite}
\usepackage{amsmath,amssymb,amsfonts}
\usepackage{amsthm}
\usepackage{graphicx}
\usepackage{textcomp}
\usepackage{algorithm}
\usepackage{algpseudocode}
\usepackage{xcolor}
\usepackage[bookmarks=false]{hyperref}
\theoremstyle{definition}
\newtheorem{definition}{Definition}
\newtheorem{theorem}{Theorem}

\DeclareMathOperator*{\argmax}{argmax}

\def\BibTeX{{\rm B\kern-.05em{\sc i\kern-.025em b}\kern-.08em
    T\kern-.1667em\lower.7ex\hbox{E}\kern-.125emX}}
    
\begin{document}

\title{DisC: Incentive Design for Green, Robust, and Fair Client Self-Participation in OTA Federated Learning}

\author{Driss Choukri$^{\bullet\circ}$, Essaid Sabir$^{\bullet}$ and Abdelkrim Haqiq$^{\circ}$\\ 
$^{\bullet}$Department of Science and Technology, TÉLUQ, University of Quebec, Montreal, H2S 3L4, Canada\\
$^{\circ}$Computer Networks, Mobility and Modeling Laboratory, FST, Hassan I University of Settat, Morocco\\
Emails: driss.choukri.doc@uhp.ac.ma, essaid.sabir@teluq.ca, abdelkrim.haqiq@uhp.ac.ma}


\maketitle

\begin{abstract}
Over-the-Air Federated Learning (OTA-FL) improves communication efficiency by exploiting waveform superposition over wireless Multiple-Access Channels (MAC). However, client heterogeneity, fading, energy constraints, and participation imbalance can degrade both fairness and aggregation accuracy. Existing Pareto-oriented fair OTA-FL schemes may still suffer from large aggregation distortion when highly weighted clients experience poor channel conditions. This paper proposes DisC, a game-theoretic distributed client selection framework for fairness-aware, channel-aware, and energy-efficient OTA-FL. Each client selects its participation level by maximizing a utility that captures Pareto fairness importance, learning relevance, OTA distortion, energy consumption, and delay. We show that DisC admits a Nash equilibrium and can be formulated as an exact potential game, where stable equilibria correspond to locally optimal OTA-FL operating points. DisC therefore offers a distributed, incentive-compatible alternative to centralized client selection under wireless and resource constraints.\vspace{.2cm}
\end{abstract}


\section{Introduction}
\subsection{Over-the-Air Federated Learning}

Federated Learning (FL) enables wireless edge devices, such as sensors, vehicles, and smartphones, to collaboratively train models for applications such as classification and prediction without sharing their raw data. OTA-FL further improves communication efficiency by exploiting MAC waveform superposition to aggregate simultaneous local updates. However, OTA-FL remains challenged by non-independent and identically distributed (non-IID) data, wireless fading, additive noise, energy limitations, and fairness imbalance. Therefore, client selection is crucial for balancing learning performance, fairness, communication reliability, and resource efficiency. Existing client selection methods include FedAvg with partial participation~\cite{mcmahan2017communication}, random selection~\cite{cho2022biased}, centralized optimization, reinforcement learning, heuristics, and meta-heuristic approaches such as the Grey Wolf Optimizer (GWO)~\cite{mirjalili2014grey,benDriss2024greenGWO}. Game-theoretic mechanisms have also been investigated for FL client selection and incentive design, including matching games~\cite{wehbi2022bilateral}, Stackelberg games~\cite{zhao2024economic}, coalition games~\cite{hu2024game}, and Nash-bargaining approaches~\cite{ni2026nash}. Nevertheless, most existing methods rely on centralized scheduling or do not explicitly capture the interaction between Pareto fairness weights and OTA channel-induced aggregation distortion. Recent fair OTA-FL schemes use multi-objective optimization and adaptive Pareto weights to improve fairness. However, OTA-FL performance strongly depends on the clients participating in each round, as they differ in data distributions, learning relevance, energy resources, and channel conditions. Selecting informative clients can improve learning, while repeatedly excluding others may reduce fairness. Conversely, clients with weak channels or limited energy may increase aggregation distortion and energy consumption. Since involving all clients in every round is often inefficient, client selection is essential for balancing learning performance, fairness, wireless reliability, and resource efficiency.

\subsection{Our Contributions}
This work develops DisC, a distributed game-theoretic client selection framework for OTA-FL that jointly accounts for fairness, learning relevance, wireless channel conditions, participation balance, energy consumption, and delay. Each client strategically selects its participation level according to a utility function whose score incorporates the distortion factor $(\bar{\lambda}_{t,k})^2/|h_{t,k}|^2$, capturing the interaction between Pareto fairness and channel quality. We prove that the resulting game admits at least one Nash equilibrium and is an exact potential game. Finally, we derive a closed-form distributed best-response rule for stable and resource-efficient client selection with limited coordination overhead.
\vspace{.2cm}

The remainder of this paper is organized as follows: Section~II presents the OTA-FL system model, including local objectives, Pareto-weighted aggregation, and OTA distortion. Section~III formulates DisC as a non-cooperative client selection game and analyzes its Nash equilibrium and potential-game structure. Section~IV derives the distributed best-response procedure and reports numerical results. Section~V concludes the paper and outlines future research directions.

\section{System Model}

We consider an OTA-FL system with one base station (BS), acting as the parameter server, and a set of wireless clients $\mathcal{N}=\{1,2,\ldots,N\}$. Each client $k\in\mathcal{N}$ owns a private dataset $\mathcal{D}_k$ and computes a local gradient vector $\mathbf{g}_{t,k}\in\mathbb{R}^{d}$ at communication round $t$, after receiving the current global model $\boldsymbol{\theta}_t$ from the BS. Due to wireless resource and energy constraints, only a subset of clients $S_t\subseteq\mathcal{N}$ participates in each round. The selected clients transmit their encoded local updates simultaneously over the wireless multiple-access channel, enabling over-the-air aggregation at the BS. The objective is to select $S_t$ by jointly accounting for learning relevance, fairness importance, channel conditions, energy consumption, delay, and OTA aggregation distortion.

\subsection{Local Objective Functions}
Each client $k\in\mathcal{N}$ minimizes its local empirical risk
$f_k(\boldsymbol{\theta})
=
\frac{1}{|\mathcal{D}_k|}
\sum_{(\mathbf{u},v)\in\mathcal{D}_k}
\ell(\mathbf{u},v|\boldsymbol{\theta})$, where $\boldsymbol{\theta}\in\mathbb{R}^{d}$ is the global model parameter, $\ell(\cdot)$ is the loss function, and $|\mathcal{D}_k|$ is the size of the local dataset. Due to data heterogeneity, minimizing only the average loss may lead to unfair performance across clients. Therefore, the FL problem is formulated as the multi-objective minimization $\min_{\boldsymbol{\theta}}
\mathbf{f}(\boldsymbol{\theta})=
\min_{\boldsymbol{\theta}}[f_1(\boldsymbol{\theta}),f_2(\boldsymbol{\theta}),\ldots,f_N(\boldsymbol{\theta})]^T$. To guide aggregation toward a fair trade-off, the BS computes adaptive Pareto weights using the modified Chebyshev method introduced in~\cite{MohajerHamidi2025}. At round $t$, the original Pareto weight vector is obtained as
\begin{equation}
\bar{\boldsymbol{\lambda}}_{t}
=
\argmax_{\boldsymbol{\lambda}\in\Delta_N}\,
\boldsymbol{\lambda}^{T}\left(\mathbf{f}(\boldsymbol{\theta}_t)-\boldsymbol{\zeta}\right)
\;\;\text{s.t.}\;\;
\left\|\boldsymbol{\lambda}-\boldsymbol{\lambda}^{\mathrm{avg}}\right\|_{\infty}\leq \epsilon,
\label{eq:original_pareto_weight}
\end{equation}
where $\Delta_N$ is the $N$-dimensional probability simplex, $\boldsymbol{\zeta}$ is a reference vector, and $\epsilon\in[0,1]$ controls the trade-off between FedAvg-like aggregation and fairness-oriented Chebyshev aggregation. The FedAvg weight of client $k$ is $\lambda_{k}^{\mathrm{avg}}=\frac{|\mathcal{D}_k|}{\sum_{j=1}^{N}|\mathcal{D}_j|}$. Since OTA aggregation is performed only over the selected set $S_t$, the original Pareto weights computed over all clients are renormalized over the active clients as
\begin{equation}
\lambda^{\star}_{t,k}
=
\frac{\bar{\lambda}_{t,k}}
{\sum_{j\in S_t}\bar{\lambda}_{t,j}+\epsilon_\lambda},
\qquad k\in S_t,
\label{eq:selected_normalized_pareto_weight}
\end{equation}
where $\epsilon_\lambda>0$ avoids numerical instability. Hence, $\sum_{k\in S_t}\lambda^{\star}_{t,k}\approx 1$. Here, $\bar{\lambda}_{t,k}$ represents the original fairness importance of client $k$ computed over all clients, while $\lambda_{t,k}^{\star}$ denotes its selected-client normalized Pareto weight used in OTA aggregation.

\subsection{Over-the-Air Aggregation Model}
At communication round $t$, a subset of clients $S_t\subseteq\mathcal{N}$ is selected for uplink transmission. The selected clients simultaneously transmit their encoded local updates over a fading Gaussian MAC, and the BS receives the superposed signal $\mathbf{y}_t=\sum_{k\in S_t}
h_{t,k}\mathbf{x}_{t,k}+\mathbf{n}_t$, where $h_{t,k}\in\mathbb{C}$ is the complex channel coefficient between client $k$ and the BS, $\mathbf{x}_{t,k}\in\mathbb{C}^{d}$ is the transmitted signal vector, and $\mathbf{n}_t\sim\mathcal{CN}(\mathbf{0},\sigma^2\mathbf{I}_d)$ is additive complex Gaussian noise. We assume a block-fading channel, where $h_{t,k}$ remains constant during one communication round and varies across rounds. Each client satisfies the average per-symbol power constraint
\begin{equation}
\mathbb{E}\left[|x_{t,k}^{(i)}|^2\right]\leq P_0,
\qquad i=1,\ldots,d,
\label{eq:power_constraint}
\end{equation}
where $P_0$ is the maximum transmit power. Although OTA aggregation reduces communication latency by allowing simultaneous transmissions, fading and noise make client selection and transmit scaling essential for reliable and fair aggregation.

\subsection{Gradient Normalization and Encoding}

Since local gradients may have heterogeneous magnitudes, each selected client computes its mean $m_{t,k}$ and variance $v_{t,k}$ and sends these two scalars over a reliable coded control channel; hence, their transmission errors are neglected. The BS then computes the Pareto-weighted global statistics as $m_t=\sum_{k\in S_t}\lambda_{t,k}^{\star}m_{t,k},\text{ and } v_t=\sum_{k\in S_t}\lambda_{t,k}^{\star}v_{t,k}$, respectively. After receiving $(m_t,v_t)$ from the BS, each selected client normalizes its local gradient as $\mathbf{s}_{t,k}=\frac{\mathbf{g}_{t,k}-m_t\mathbf{1}_d}
{\sqrt{v_t+\epsilon_v}}$, where $\epsilon_v>0$ ensures numerical stability. The normalized vector is then linearly encoded as $\mathbf{x}_{t,k}=b_{t,k}\mathbf{s}_{t,k}$, where $b_{t,k}\in\mathbb{C}$ is the transmit scalar chosen to satisfy the power constraint in \eqref{eq:power_constraint} and to enable unbiased Pareto-weighted OTA aggregation.

\subsection{OTA Decoding and Aggregation Error}

After receiving the superposed signal $\mathbf{y}_t$, the BS applies a receive de-noising scalar $c_t$ and estimates the aggregated gradient as $\hat{\mathbf{g}}_t=\frac{\sqrt{v_t+\epsilon_v}}{c_t}\mathbf{y}_t+m_t\mathbf{1}_d$. The desired Pareto-weighted gradient over the selected clients is $\mathbf{g}_t^{\lambda} =\sum_{k\in S_t}\lambda_{t,k}^{\star}\mathbf{g}_{t,k}$. To obtain an unbiased estimate of $\mathbf{g}_t^{\lambda}$, the transmit and receive scalars are chosen as $b_{t,k}=\frac{\lambda_{t,k}^{\star}c_t}{h_{t,k}}, \text{ and }c_t=\min_{k\in S_t}\frac{\sqrt{P_0}|h_{t,k}|}{\lambda_{t,k}^{\star}}$, respectively. With these choices, $\mathbb{E}[\hat{\mathbf{g}}_t]=\mathbf{g}_t^{\lambda}$. The OTA aggregation distortion is defined as $\mathcal{E}_t=\mathbb{E}\left[
\left\|\hat{\mathbf{g}}_t-\mathbf{g}_t^{\lambda}\right\|_2^2\right]$, and is proportional to
\begin{equation}
\mathcal{E}_t
\propto
\frac{d(v_t+\epsilon_v)\sigma^2}{P_0}
\max_{k\in S_t}
\frac{(\lambda_{t,k}^{\star})^2}
{|h_{t,k}|^2+\epsilon_h},
\label{eq:ota_error}
\end{equation}
where $\epsilon_h>0$ avoids numerical instability for very small channel gains. Equation~\eqref{eq:ota_error} shows the fairness--communication trade-off in OTA-FL: a client with a large Pareto weight improves fairness-oriented learning, but if it has a weak channel, it may significantly increase the OTA aggregation distortion. This motivates the proposed fairness-aware and channel-aware client selection game.

\section{Client Selection as a Non-Cooperative Game}

\subsection{The Game Structure}
Based on the fairness--communication trade-off identified in \eqref{eq:ota_error}, we formulate the client selection problem as a non-cooperative game among wireless clients. This formulation is motivated by recent works showing that strategic decision-making can improve client selection, incentives, participation stability, and resource allocation in FL systems~\cite{wehbi2022bilateral,zhao2024economic,hu2024game,ni2026nash}. The objective is to enable each client to strategically determine its participation level by considering its fairness importance, learning contribution, channel condition, energy cost, delay, and impact on OTA aggregation distortion. The DisC game is defined as $\mathcal{G}_t=\left(\mathcal{N},
\{\mathcal{A}_{t,k}\}_{k\in\mathcal{N}},\{U_{t,k}\}_{k\in\mathcal{N}}
\right)$, where $\mathcal{N}$ denotes the set of players, $\mathcal{A}_{t,k}$ is the strategy set of client $k$ at communication round $t$, and $U_{t,k}$ is the utility function of client $k$. The utility function is designed to capture both the individual benefit of participation and the global impact of the client on OTA-FL performance. In this formulation, clients are modeled as rational decision-makers that seek to maximize their own utilities. A stable client selection configuration is obtained when no client can improve its utility by unilaterally changing its participation decision. Such a stable point corresponds to a Nash equilibrium of the DisC client selection game.\vspace{.2cm}

\subsubsection{The set of players} 
The players of the DisC game are the wireless FL clients in the set $\mathcal{N}$. Each client $k\in\mathcal{N}$ is modeled as a rational and autonomous decision-maker that chooses its participation behavior at communication round $t$. In practical OTA-FL systems, clients may have different local data distributions, battery levels, computation capabilities, channel conditions, and delay constraints. Therefore, their willingness and ability to participate may vary from one round to another. At each round, client $k$ decides whether to participate in the OTA aggregation process and, in the relaxed formulation, how strongly it participates. This participation decision affects not only the client's own utility, through energy consumption and delay, but also the global OTA-FL performance, through its learning contribution, fairness importance, and impact on aggregation distortion.\vspace{.2cm}

\subsubsection{The Strategy Space}
At communication round $t$, each client $k\in\mathcal{N}$ chooses a participation level $a_{t,k}\in[0,1]$, where $a_{t,k}=0$ indicates that client $k$ does not participate, while $a_{t,k}=1$ indicates full participation in the current OTA aggregation round. Values between $0$ and $1$ can be interpreted either as relaxed participation decisions or as participation probabilities. Accordingly, the strategy set of client $k$ is defined as $\mathcal{A}_{t,k}=[0,1]$. The joint strategy profile of all clients at round $t$ is given by $a_t=(a_{t,1},a_{t,2},\ldots,a_{t,N})\in\mathcal{A}_t$, where $\mathcal{A}_t=\prod_{k\in\mathcal{N}}\mathcal{A}_{t,k}$ denotes the joint strategy space. The continuous strategy formulation provides a tractable relaxation of the binary client selection problem. After the game converges to a stable participation profile $a_t^{\star}$, the BS constructs the scheduled client set $S_t$ by selecting the $M$ clients with the largest participation levels:
\begin{equation}
S_t
=
\operatorname{TopM}
\left(
a_{t,1}^{\star},
a_{t,2}^{\star},
\ldots,
a_{t,N}^{\star}
\right),
\label{eq:topM_selection}
\end{equation}
where $M$ denotes the maximum number of clients allowed to participate in one communication round. This mapping converts the relaxed game outcome into a feasible OTA-FL scheduling decision.\vspace{.2cm}

\subsubsection{Fairness-Aware and Channel-Aware Client Score}
Before defining the utility function, we introduce an individual selection score that quantifies the overall contribution of each client to the OTA-FL process. For each client $k$ at communication round $t$, we first define the normalized quantities used in the score. The term $\widetilde{\bar{\lambda}}_{t,k}$ denotes the normalized Pareto fairness weight of client $k$, where a larger value indicates higher importance in fairness-oriented aggregation before the final OTA scheduling step. The term $\widetilde{q}_{t,k}$ captures the normalized learning relevance of client $k$, which may be measured using its local loss, gradient norm, data quality, or update significance. The term $\widetilde{A}_{t,k}$ represents the normalized participation debt, also referred to as the age-of-participation, which promotes long-term participation fairness by giving higher priority to clients that have not been selected for several rounds~\cite{Liu2026}. In contrast, $\widetilde{E}_{t,k}$ and $\widetilde{D}_{t,k}$ denote the normalized energy consumption and delay, respectively, and are therefore included as penalty terms. The term $\widetilde{\Psi}_{t,k}$ represents the normalized OTA distortion factor of client $k$. It captures the joint effect of the Pareto fairness weight and the wireless channel gain on the aggregation error. Specifically, we define the unnormalized distortion factor as $\Psi_{t,k}=\frac{(\bar{\lambda}_{t,k})^2}{|h_{t,k}|^2+\epsilon_h}$,
where $\epsilon_h>0$ is a small constant introduced to avoid numerical instability when the channel gain is very small. This factor is directly motivated by the OTA aggregation error in \eqref{eq:ota_error}. Therefore, including $\widetilde{\Psi}_{t,k}$ in the client score discourages the selection of clients that may significantly increase OTA aggregation distortion, unless their learning and fairness contributions are sufficiently important. Using these normalized quantities, the score of client $k$ at communication round $t$ is defined as
\begin{equation}
R_{t,k}
=
\alpha \widetilde{\bar{\lambda}}_{t,k}
+
\beta \widetilde{q}_{t,k}
+
\zeta \widetilde{A}_{t,k}
-
\gamma \widetilde{\Psi}_{t,k}
-
\delta \widetilde{E}_{t,k}
-
\mu \widetilde{D}_{t,k},
\label{eq:client_score}
\end{equation}
where $\alpha,\beta,\zeta,\gamma,\delta,\mu\geq0$ are design parameters that control the relative importance of fairness, learning relevance, participation fairness, OTA distortion, energy consumption, and delay, respectively. To ensure that all attributes are comparable, each metric is normalized before being combined in \eqref{eq:client_score}. For a generic positive metric $X_{t,k}$, its normalized value can be computed as
\begin{equation}
\widetilde{X}_{t,k}
=
\frac{X_{t,k}-X_t^{\min}}
{X_t^{\max}-X_t^{\min}+\epsilon_x},
\label{eq:normalization}
\end{equation}
where $X_t^{\min}$ and $X_t^{\max}$ are the minimum and maximum values of the metric among all clients at round $t$, and $\epsilon_x>0$ avoids division by zero. The coefficients are selected empirically through validation and sensitivity analysis. At each round, the BS updates the fairness weights and participation debt from training history, while clients measure their local learning relevance, channel quality, energy, and delay; min--max normalization then makes these metrics comparable.\vspace{.2cm}

\subsubsection{Utility Function}
Based on the client score defined in \eqref{eq:client_score}, we define the utility function of client $k$ at communication round $t$ as
\begin{equation}
U_{t,k}(a_{t,k},a_{t,-k})
=
a_{t,k}R_{t,k}
-
\frac{\kappa}{2}
\left(
\sum_{j\in\mathcal{N}}a_{t,j}-M
\right)^2
-
\frac{\nu}{2}a_{t,k}^2,
\label{eq:proposed_utility}
\end{equation}
where $a_{t,-k}$ denotes the participation profile of all clients except client $k$, $M$ is the maximum desired number of selected clients, and $\kappa,\nu>0$ are regularization parameters. The first term, $a_{t,k}R_{t,k}$, represents the individual participation benefit of client $k$. A client with a high score $R_{t,k}$ obtains a larger utility from participating in the current OTA-FL round. This encourages the participation of clients that provide a favorable trade-off among fairness importance, learning relevance, participation debt, OTA aggregation reliability, energy consumption, and delay. The second term penalizes deviations from the desired number of participating clients. It introduces coupling among clients because the utility of each client depends on the aggregate participation level $\sum_{j\in\mathcal{N}}a_{t,j}$. This term prevents excessive participation, which may increase OTA aggregation distortion and resource consumption, while also avoiding insufficient participation, which may slow down the learning process. The third term, $-\frac{\nu}{2}a_{t,k}^2$, is a participation regularization term. It discourages overly aggressive participation decisions and ensures that the utility function is strictly concave with respect to the client's own strategy $a_{t,k}$. This property is useful for establishing the existence of a stable equilibrium and deriving a well-defined best-response update.

Therefore, maximizing \eqref{eq:proposed_utility} enables each client to strategically determine its participation level by balancing its individual contribution to fairness-aware learning with the global constraints imposed by wireless resources and OTA aggregation reliability.

\subsection{Nash Equilibrium Analysis}

In this section, we analyze the stability of the DisC client selection game. Since each client strategically chooses its participation level to maximize its own utility, a stable operating point is reached when no client has an incentive to change its participation decision unilaterally. This stable point is characterized by the notion of Nash equilibrium.

\subsubsection{Nash Equilibrium Definition}

\begin{definition}[Nash Equilibrium]
A strategy profile $a_t^{\star}
=
(a_{t,1}^{\star},a_{t,2}^{\star},\ldots,a_{t,N}^{\star})$
is a Nash equilibrium of the DisC client selection game if no client can improve its utility by unilaterally changing its participation level while the participation levels of all other clients remain fixed. That is, for every client $k\in\mathcal{N}$,
\begin{equation}
U_{t,k}(a_{t,k}^{\star},a_{t,-k}^{\star})
\geq
U_{t,k}(a_{t,k},a_{t,-k}^{\star}),
\qquad
\forall a_{t,k}\in\mathcal{A}_{t,k}.
\label{eq:nash_equilibrium}
\end{equation}
\end{definition}

At a Nash equilibrium, each client’s participation level is a best response to the participation decisions of the other clients. Therefore, the equilibrium profile $a_t^{\star}$ represents a stable client selection configuration for the current OTA-FL round. After obtaining $a_t^{\star}$, the BS can construct the scheduled client set by selecting the clients with the largest equilibrium participation levels.

\subsubsection{Existence of Nash Equilibrium}

We next establish that the DisC client selection game admits at least one stable operating point. The existence result is important because it guarantees that the strategic interaction among clients can converge to a participation profile in which no client has an incentive to deviate unilaterally.

\begin{theorem}
The DisC OTA-FL client selection game admits at least one Nash equilibrium.
\end{theorem}

\begin{proof}
For each client $k\in\mathcal{N}$, the strategy set is given by
$\mathcal{A}_{t,k}=[0,1]$. This set is non-empty, compact, and convex. Hence, the joint strategy space $\mathcal{A}_t=\prod_{k\in\mathcal{N}}\mathcal{A}_{t,k}$ is also non-empty, compact, and convex. The utility function of client $k$ is given by \eqref{eq:proposed_utility}. Since $U_{t,k}$ is composed of linear and quadratic terms in the joint strategy profile $a_t$, it is continuous in $a_t$. Furthermore, for a fixed $a_{t,-k}$, the second derivative of $U_{t,k}$ with respect to $a_{t,k}$ is $\frac{\partial^2 U_{t,k}}{\partial a_{t,k}^2}=-\kappa-\nu < 0$, because $\kappa>0$ and $\nu>0$. Therefore, $U_{t,k}$ is strictly concave in the own strategy $a_{t,k}$ of client $k$. Thus, the DisC game satisfies the conditions of the Debreu--Glicksberg--Fan theorem \cite{Debreu1952,Glicksberg1952}. Consequently, the DisC OTA-FL client selection game admits at least one Nash equilibrium.
\end{proof}

\subsection{Potential Game Interpretation}
To link individual client decisions with the global OTA-FL selection objective, we interpret the DisC game as a potential game. Define the potential function as
\begin{equation}
\Phi_t(a_t)
=
\sum_{k\in\mathcal{N}}
a_{t,k}R_{t,k}
-
\frac{\kappa}{2}
\left(
\sum_{k\in\mathcal{N}}a_{t,k}-M
\right)^2
-
\frac{\nu}{2}
\sum_{k\in\mathcal{N}}a_{t,k}^2.
\label{eq:potential_function}
\end{equation}
The first term represents the total benefit of client participation, while the second and third terms penalize excessive or insufficient participation and regularize the participation levels.

For any unilateral deviation of client $k$ from $a_{t,k}$ to $a'_{t,k}$, with $a_{t,-k}$ fixed, the variation in the utility function is equal to the variation in the potential function $U_{t,k}(a'_{t,k},a_{t,-k})-U_{t,k}(a_{t,k},a_{t,-k}) =\Phi_t(a'_{t,k},a_{t,-k})-\Phi_t(a_{t,k},a_{t,-k})$. Hence, the DisC client selection game is an exact potential game. Therefore, every local maximizer of $\Phi_t(a_t)$ corresponds to a Nash equilibrium. This shows that decentralized participation decisions are aligned with the global objective of selecting clients that jointly improve fairness, learning relevance, OTA aggregation reliability, energy efficiency, and delay performance.

\section{Distributed Near-Optimal Client Selection}

\subsection{Distributed Best-Response Update}

The potential-game structure allows clients to reach a stable participation profile through distributed best-response updates. At iteration $r$, client $k$ updates its participation level by solving
\begin{equation}
a_{t,k}^{(r+1)}
=
\arg\max_{a_{t,k}\in[0,1]}
U_{t,k}\left(a_{t,k},a_{t,-k}^{(r)}\right).
\label{eq:best_response_problem}
\end{equation}
From \eqref{eq:proposed_utility}, the closed-form best response is
\begin{equation}
a_{t,k}^{(r+1)}
=
\Pi_{[0,1]}
\left[
\frac{
R_{t,k}
-
\kappa
\left(
\sum_{j\neq k}a_{t,j}^{(r)}-M
\right)
}
{\kappa+\nu}
\right],
\label{eq:best_response}
\end{equation}
where $\Pi_{[0,1]}(x)=\min\{1,\max\{0,x\}\}$ is the projection onto $[0,1]$. The update in \eqref{eq:best_response} increases the participation level of clients with high scores $R_{t,k}$, while reducing participation when the aggregate participation of other clients approaches or exceeds the target number $M$. After convergence, the BS selects the $M$ clients with the largest equilibrium participation levels $S_t^{\star}=\operatorname{TopM}
\left(a_{t,1}^{\star},a_{t,2}^{\star},\ldots,a_{t,N}^{\star}
\right)$. The selected set $S_t^{\star}$ is then used for gradient normalization, analog transmission, and OTA aggregation.

\subsection{Proposed OTA-FL Procedure}
Algorithm~\ref{alg:disc_otafl} summarizes the proposed DisC-based OTA-FL procedure.
\mbox{}\\[-1.2\baselineskip]
\begin{algorithm}[h]
\caption{DisC-Based OTA-FL Procedure}
\label{alg:disc_otafl}
\begin{algorithmic}[1]
\Require $N$, $M$, $T$, initial model $\boldsymbol{\theta}_0$
\For{$t=0,\ldots,T-1$}
    \State BS broadcasts $\boldsymbol{\theta}_t$; each client computes its local metrics.
    \State BS computes $\bar{\boldsymbol{\lambda}}_t$; each client evaluates
    $\Psi_{t,k}=\frac{(\bar{\lambda}_{t,k})^2}{|h_{t,k}|^2+\epsilon_h}$ and $R_{t,k}$.
    \State Clients update $a_{t,k}^{\star}$ using \eqref{eq:best_response}.
    \State BS selects $S_t^{\star}=\operatorname{TopM}(a_{t,1}^{\star},\ldots,a_{t,N}^{\star})$.
    \State BS normalizes the selected-client weights using \eqref{eq:selected_normalized_pareto_weight}.
    \State Selected clients send $(m_{t,k},v_{t,k})$; BS broadcasts $(m_t,v_t)$.
    \State Selected clients normalize, encode with $b_{t,k}$, and transmit over the fading Gaussian MAC.
    \State BS applies $c_t$, reconstructs $\hat{\mathbf{g}}_t$, and updates
    $\boldsymbol{\theta}_{t+1}=\boldsymbol{\theta}_t-\eta_t\hat{\mathbf{g}}_t$.
\EndFor
\Ensure Final global model $\boldsymbol{\theta}_T$
\end{algorithmic}
\end{algorithm}
\subsection{Numerical Results}

We compare DisC with FedAvg~\cite{mcmahan2017communication}, random selection~\cite{cho2022biased}, and GWO~\cite{mirjalili2014grey,benDriss2024greenGWO} using $N=20$, $M=10$, and $T=200$. Following~\cite{benDriss2024greenGWO}, transmission energy is evaluated as transmit power times transmission duration, while latency is obtained from the update size and achievable uplink rate. We report accuracy, OTA distortion, energy, participation rate, and Jain's fairness index.\vspace{.2cm}

\begin{figure}[h]
    \centering
    \includegraphics[width=0.8\columnwidth]{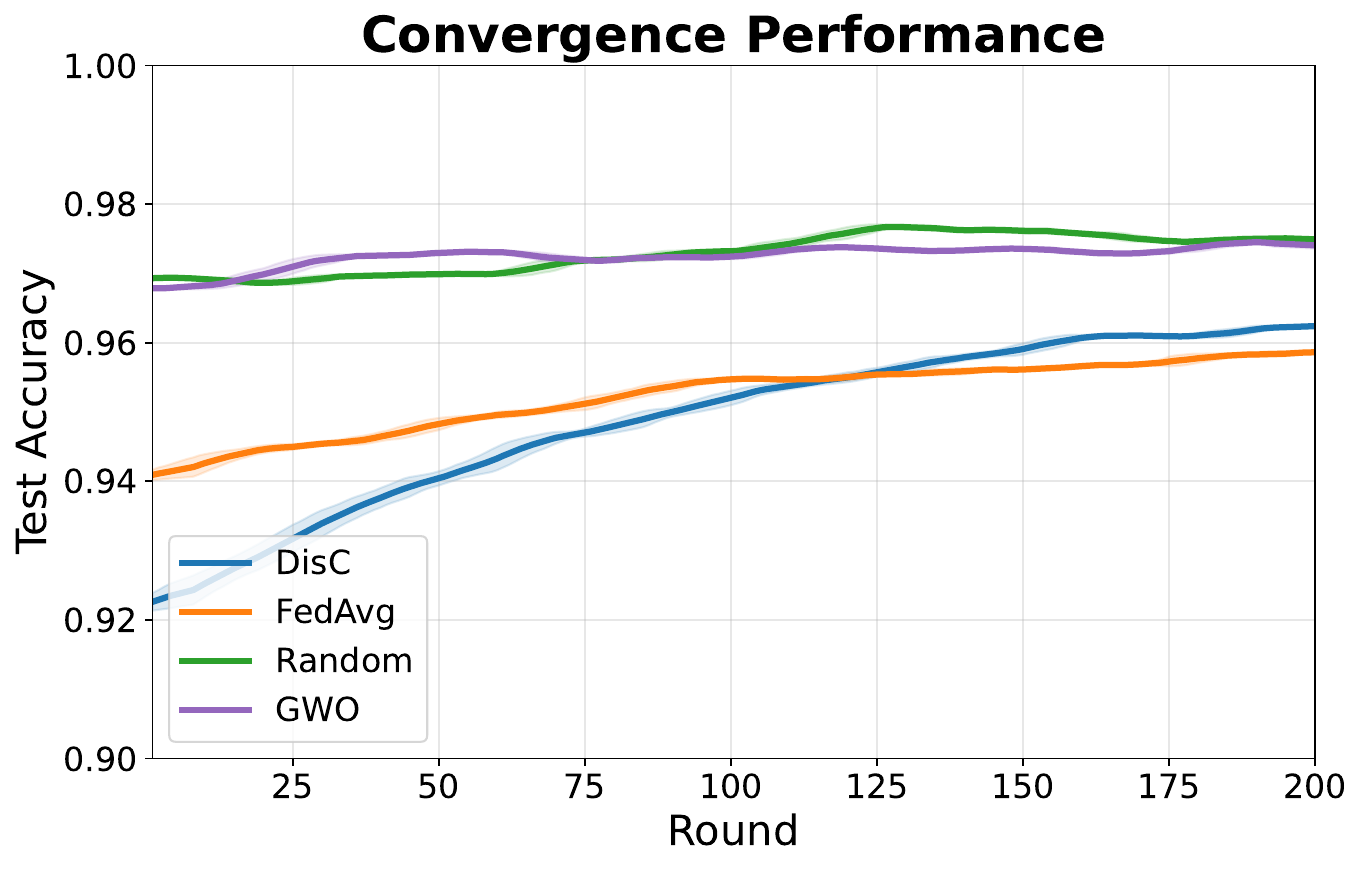}
    \caption{Convergence performance of the considered methods.}
    \label{fig:accuracy}
\end{figure}

\noindent\textbf{Convergence and accuracy:}
\hyperref[fig:accuracy]{Fig.~\ref*{fig:accuracy}} shows stable convergence for all methods. 
DisC achieves competitive accuracy compared with FedAvg, while random selection and GWO obtain slightly higher final accuracies. 
This confirms that DisC preserves learning performance while accounting for wireless reliability, energy consumption, and participation fairness. 
Its closed-form best response in \eqref{eq:best_response} requires only scalar information, yielding $\mathcal{O}(1)$ per-client cost per iteration, while the BS performs Top-$M$ selection in $\mathcal{O}(N\log N)$ time, reducible to $\mathcal{O}(N)$.

\begin{figure}[h]
\centering
\includegraphics[width=0.8\columnwidth]{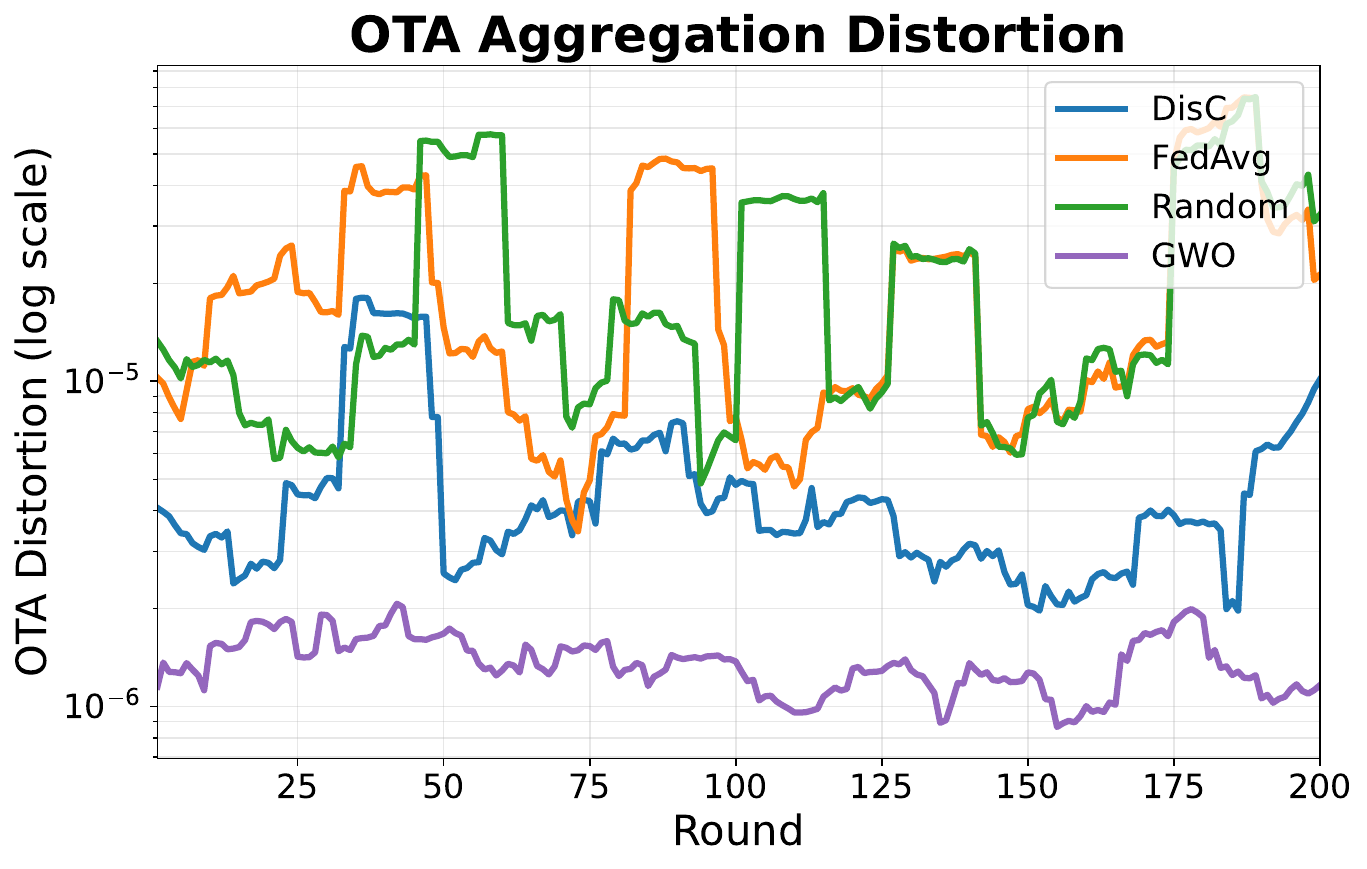}
 \caption{OTA aggregation distortion of the considered client-selection schemes.}
\label{fig:distortion}
\end{figure}

\noindent\textbf{Robustness and aggregation distortion:}
\hyperref[fig:distortion]{Fig.~\ref*{fig:distortion}} reports the OTA aggregation distortion in logarithmic scale. DisC achieves lower distortion than FedAvg and random selection, confirming the benefit of the OTA distortion-aware score. GWO obtains the lowest distortion due to its stronger preference for favorable channel conditions.

\begin{figure}[h]
    \centering
    \includegraphics[width=0.8\columnwidth]{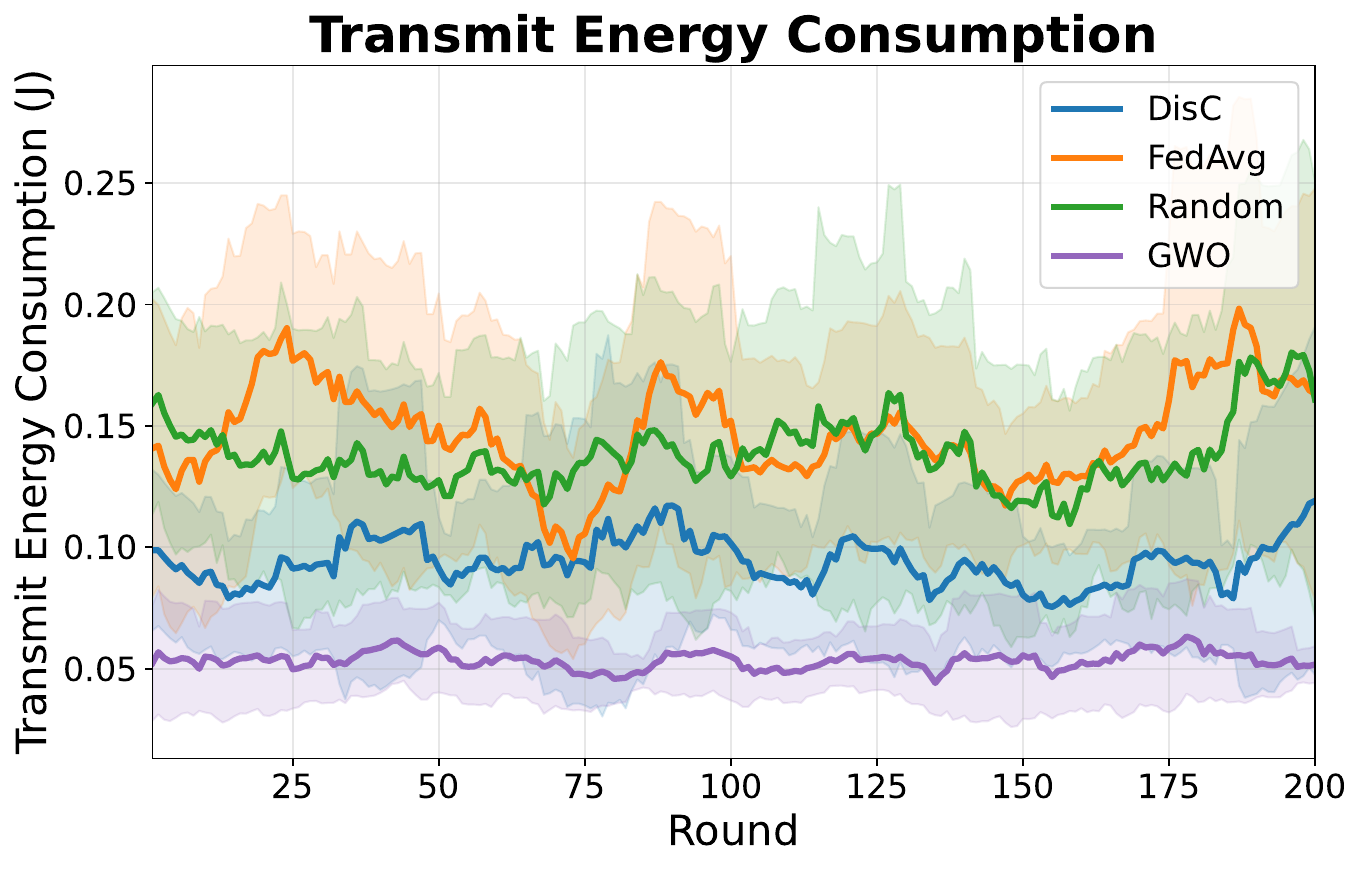}
    \caption{Transmit energy consumption of the considered methods.}
    \label{fig:energy}
\end{figure}

\noindent\textbf{Energy efficiency:}
\hyperref[fig:energy]{Fig.~\ref*{fig:energy}} compares the transmit energy consumption. DisC consumes less energy than FedAvg and random selection, while GWO remains the most energy-efficient. This shows that DisC provides a favorable trade-off between energy efficiency, OTA reliability, and fairness.

\begin{figure}[t]
    \centering
    \includegraphics[width=0.8\columnwidth]{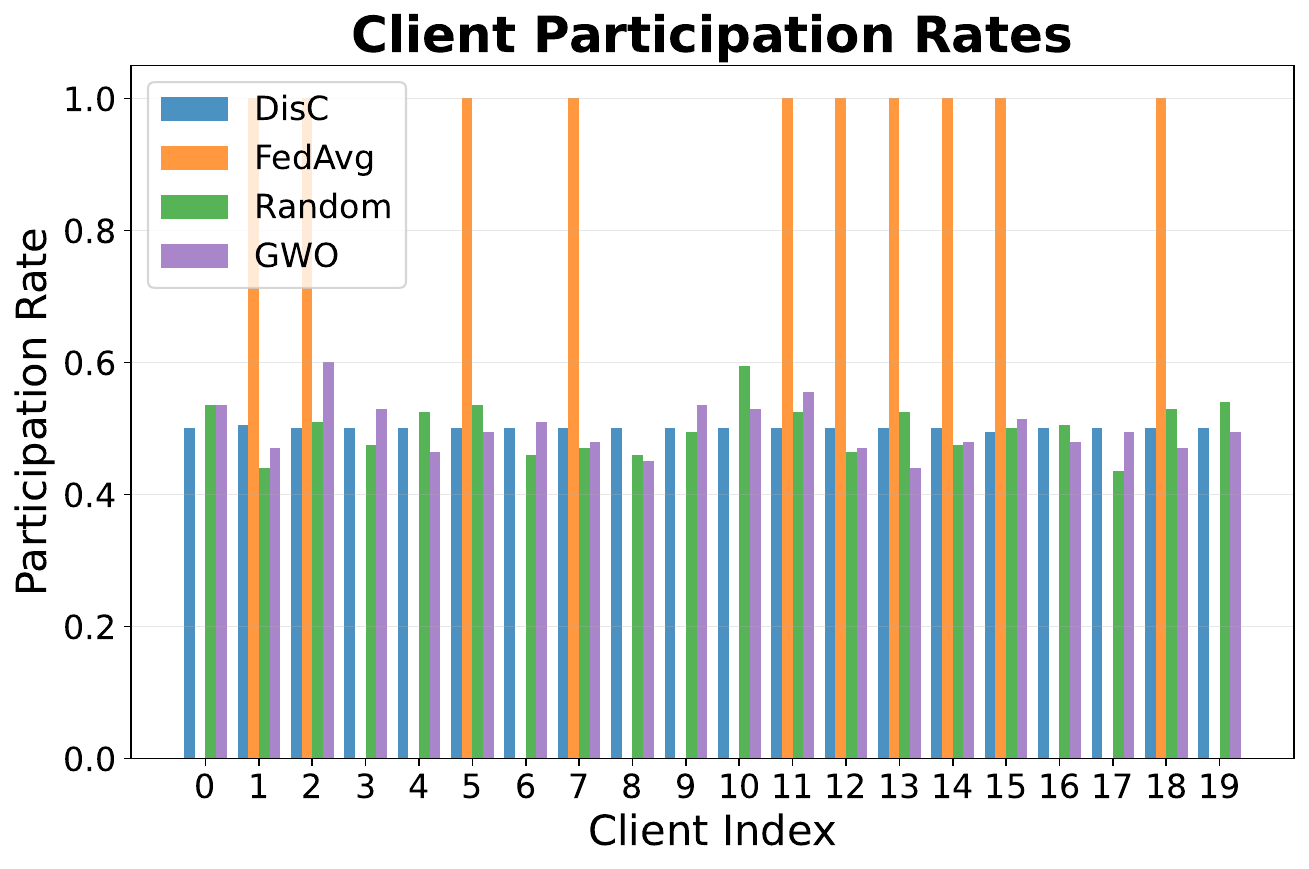}
    \caption{Long-term client participation rates.}
    \label{fig:participation}
\end{figure}

\noindent\textbf{Willingness to participate and fairness:}
\hyperref[fig:participation]{Fig.~\ref*{fig:participation}} shows the long-term client participation rates. Since $M=10$ clients are selected among $N=20$, the ideal participation rate is $M/N=0.5$. DisC keeps client participation close to this value, unlike FedAvg, which repeatedly selects a fixed subset. This confirms the role of the participation-debt term in improving long-term fairness.
The average results over the last 50 communication rounds are summarized in \hyperref[tab:ota_fl_summary]{Table~\ref*{tab:ota_fl_summary}}.
\begin{table}[h]
\centering
\caption{Summary of the average performance over the last 50 communication rounds.}
\label{tab:ota_fl_summary}
\renewcommand{\arraystretch}{1.15}
\setlength{\tabcolsep}{4pt}
\small
\begin{tabular}{lcccc}
\hline
\textbf{Method} & \textbf{Acc. $\uparrow$} & \textbf{Dist. $\downarrow$} & \textbf{Energy $\downarrow$} & \textbf{Jain $\uparrow$} \\
\hline
DisC   & 0.9612 & $3.91{\times}10^{-6}$ & 0.091 & 1.000 \\
FedAvg & 0.9574 & $3.04{\times}10^{-5}$ & 0.156 & 0.500 \\
Random & 0.9752 & $2.97{\times}10^{-5}$ & 0.138 & 0.994 \\
GWO    & 0.9736 & $1.28{\times}10^{-6}$ & 0.055 & 0.994 \\
\hline
\end{tabular}
\end{table}

\hyperref[tab:ota_fl_summary]{Table~\ref*{tab:ota_fl_summary}} confirms that DisC achieves the best participation fairness, with Jain's index equal to $1.000$. It also reduces OTA distortion and energy consumption compared with FedAvg and random selection. Although GWO achieves the lowest distortion and energy consumption, and random selection attains the highest accuracy, DisC provides the most balanced accuracy--distortion--energy--fairness trade-off.\vspace{.2cm}

\begin{figure}[h]
    \centering
    \setlength{\tabcolsep}{1pt}
    \begin{tabular}{cc}
        \includegraphics[width=0.49\columnwidth]{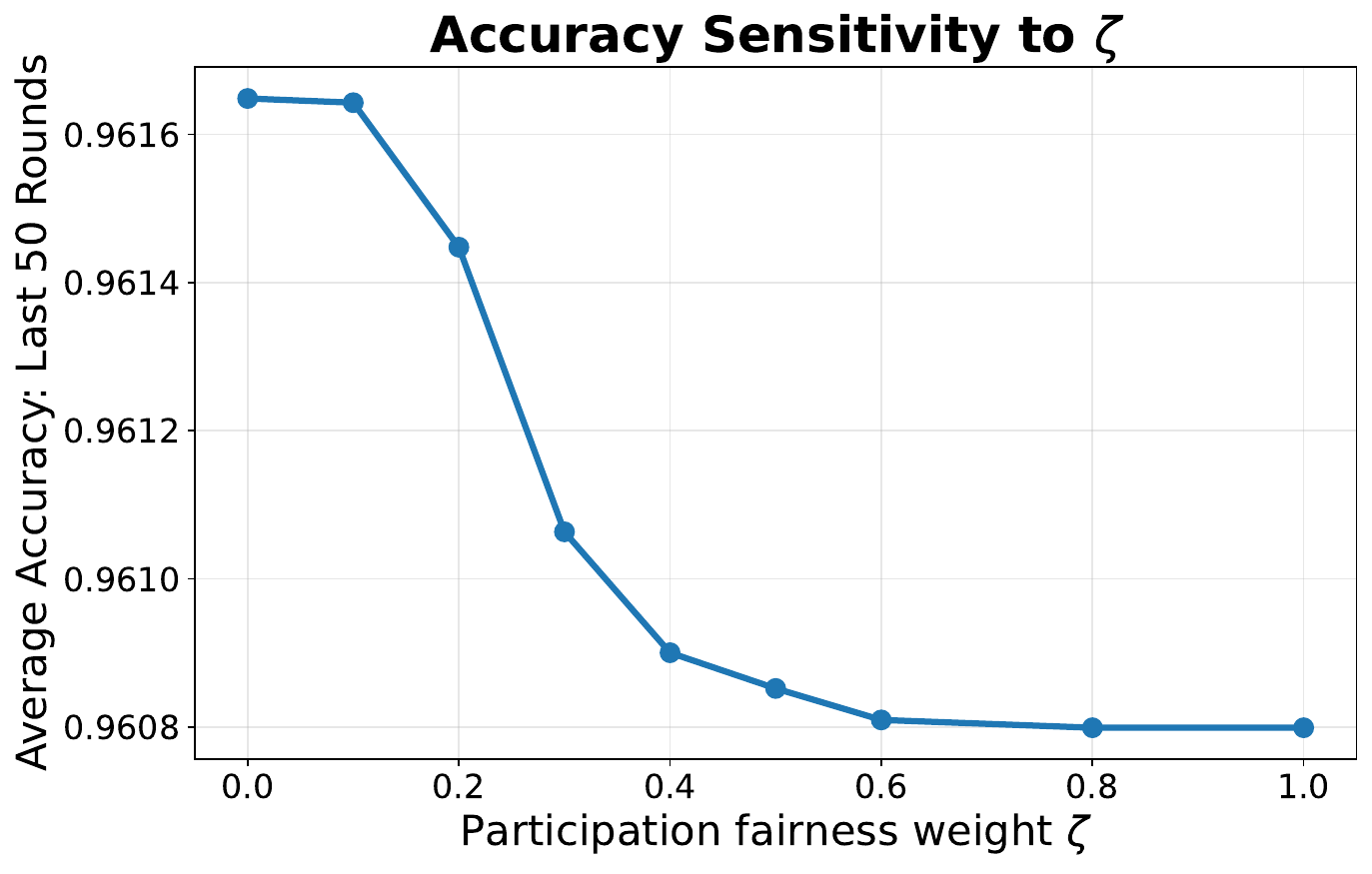}
        &
        \includegraphics[width=0.49\columnwidth]{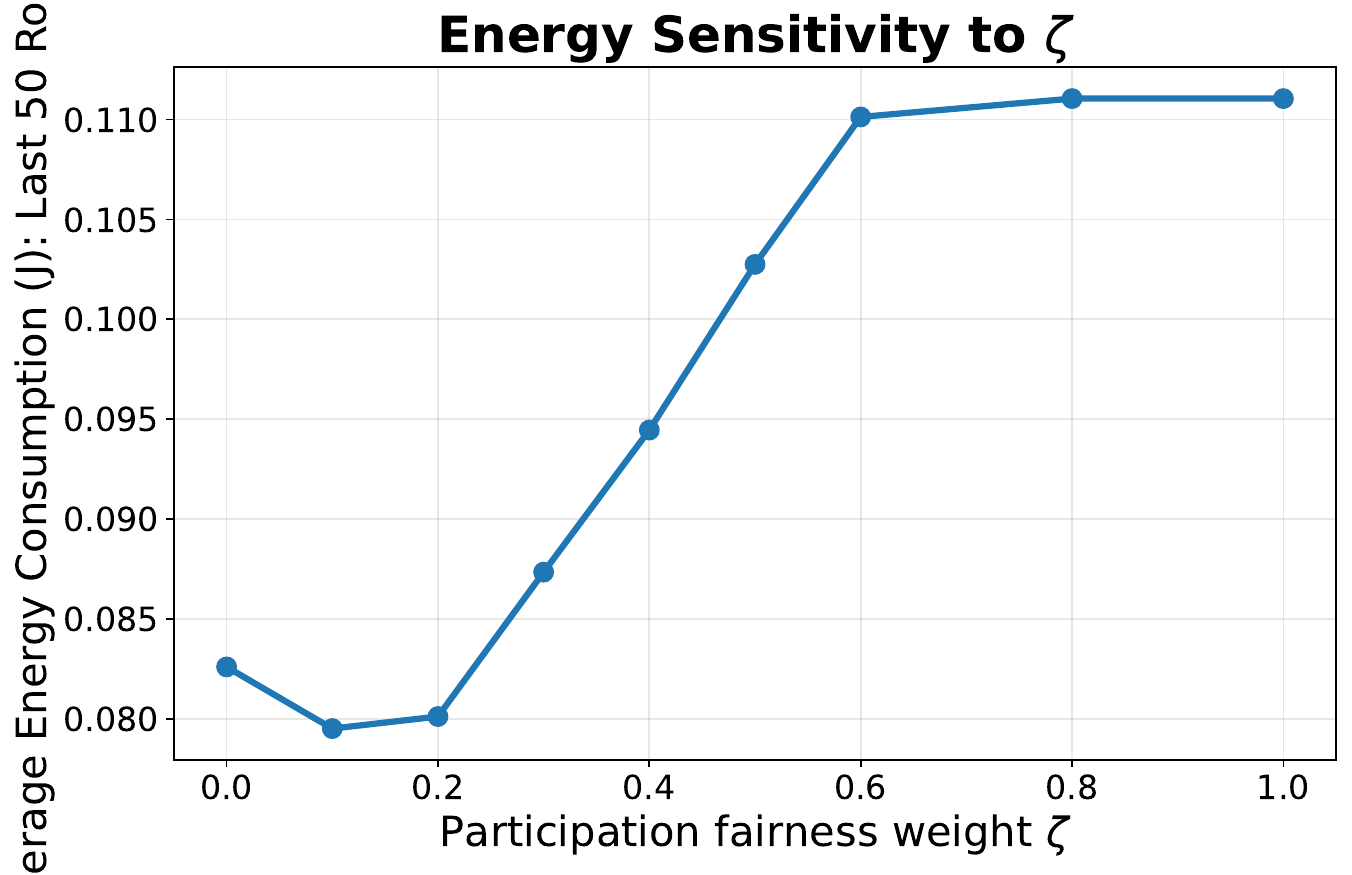}
        \\
        {\footnotesize (a) Accuracy}
        &
        {\footnotesize (b) Energy}
        \\[0.1cm]
        \includegraphics[width=0.49\columnwidth]{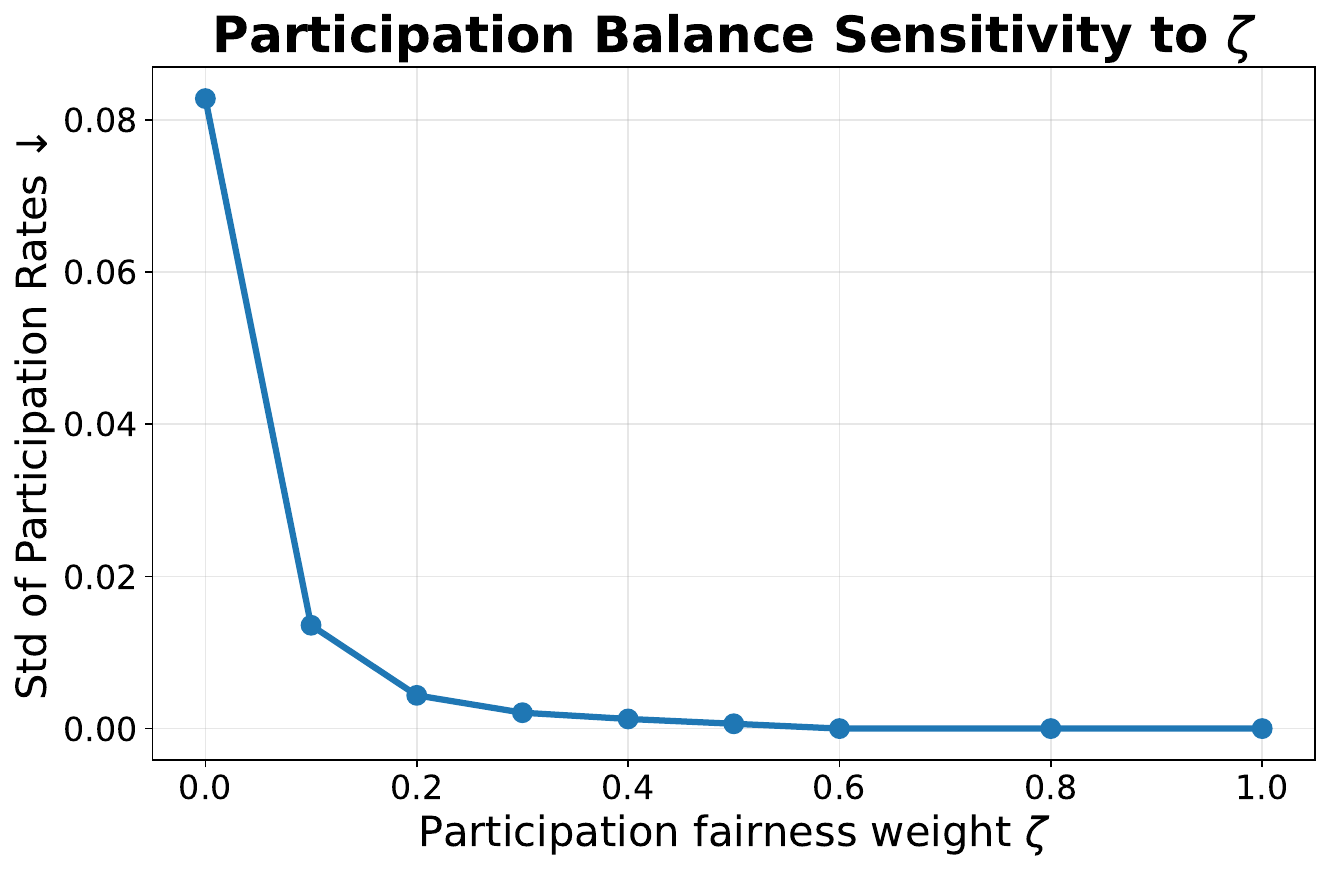}
        &
        \includegraphics[width=0.49\columnwidth]{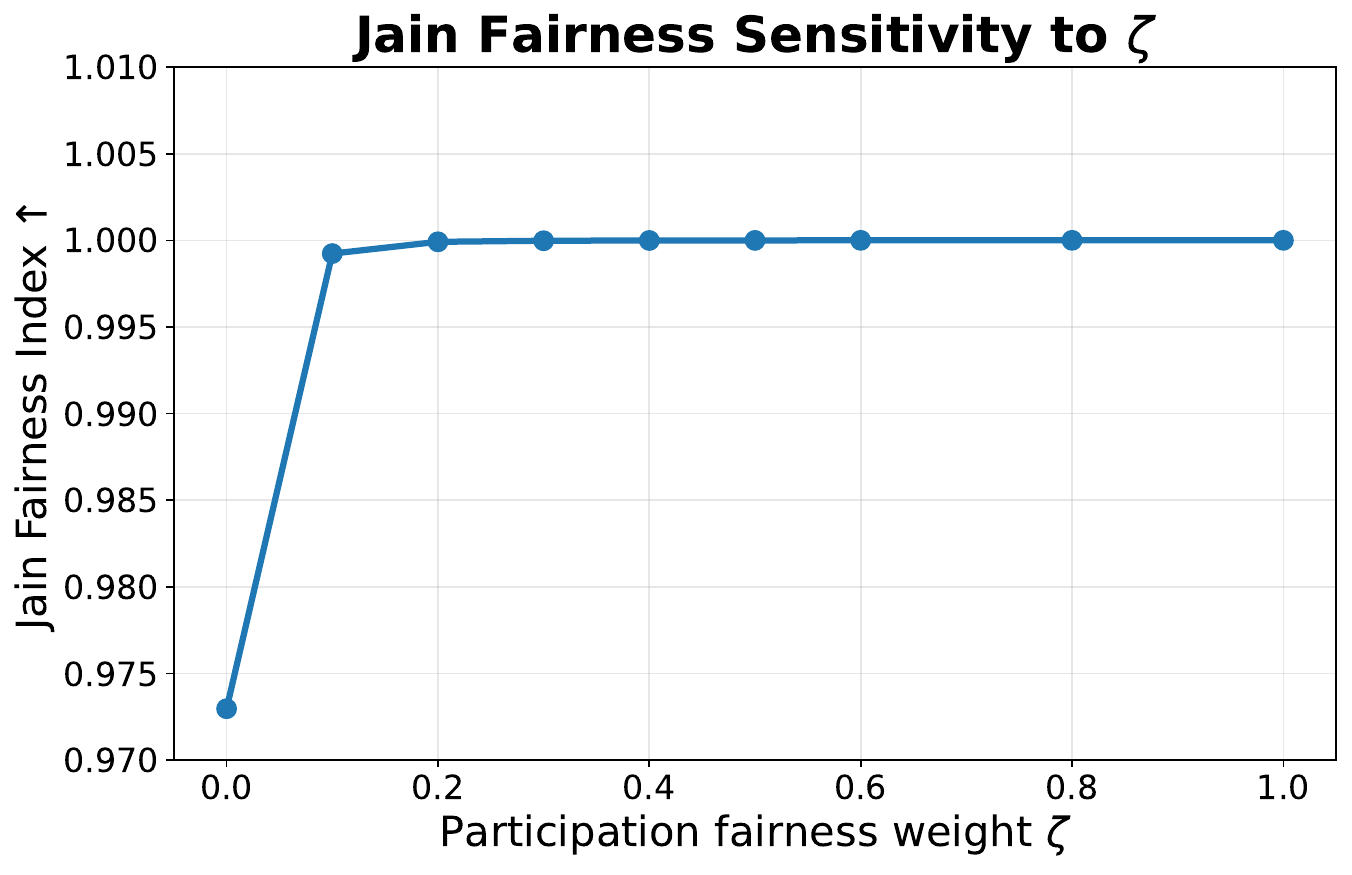}
        \\
        {\footnotesize (c) Participation std}
        &
        {\footnotesize (d) Jain index}
    \end{tabular}
    \caption{Sensitivity of DisC with respect to the participation-fairness weight~$\zeta$.}
    \label{fig:zeta_sensitivity}\vspace{-.5cm}
\end{figure}

\noindent\textbf{Sensitivity to $\zeta$:}
\hyperref[fig:zeta_sensitivity]{Fig.~\ref*{fig:zeta_sensitivity}} shows the effect of the participation-fairness weight $\zeta$. Increasing $\zeta$ improves participation balance and Jain fairness, while slightly reducing accuracy and increasing energy consumption. This confirms that $\zeta$ controls the trade-off between fairness and performance-oriented client selection.\vspace{.2cm}

\noindent\textbf{Tradeoff analysis:}
Unlike existing game-theoretic FL methods designed for pricing, matching, coalition formation, or bargaining, DisC addresses Pareto-weighted OTA aggregation and is therefore not directly comparable with them. We thus use FedAvg, random selection, and GWO as representative baselines under identical conditions. Although random selection and GWO may outperform DisC in individual metrics, DisC provides a more balanced accuracy--distortion--energy--fairness trade-off by explicitly incorporating fairness, channel reliability, energy, and delay into its utility function.

\section{Conclusion}
This paper introduces DisC, a game-theoretic framework for fair, channel-aware, and energy-efficient client selection in OTA-FL. The selection problem is formulated as a non-cooperative potential game with at least one Nash equilibrium, and a distributed best-response rule enables stable participation decisions. Numerical results confirm that DisC improves participation fairness while reducing OTA distortion and energy consumption compared with FedAvg, GWO and random selection. Future work will consider dynamic energy harvesting, imperfect channel state information, and convergence analysis under time-varying client participation.

\bibliographystyle{IEEEtran}
\bibliography{references}
\end{document}